\documentclass[a4paper,11pt]{article}
\usepackage{jcappub}
\usepackage{orcidlink}
\usepackage{cmap}
\usepackage[utf8]{inputenc}
\usepackage[T1]{fontenc}

\def\imo{i}
\def\re#1{\mathrm{Re}(#1)}
\def\im#1{\mathrm{Im}(#1)}
\def\K{{\cal K}}
\def\Order#1{{\cal O}\left(#1\right)}

\begin{document}

\title{Quasinormal ringing and shadows of black holes and wormholes in dark matter-inspired Weyl gravity}

\author[\dagger]{Roman A. Konoplya \orcidlink{0000-0003-1343-9584},}
\emailAdd{roman.konoplya@gmail.com}
\affiliation[\dagger]{Research Centre for Theoretical Physics and Astrophysics, \\ Institute of Physics, Silesian University in Opava, \\ Bezručovo náměstí 13, CZ-74601 Opava, Czech Republic}

\author[\ddagger]{Andrii Khrabustovskyi \orcidlink{0000-0001-6298-9684},}
\emailAdd{andrii.khrabustovskyi@uhk.cz}

\author[\ddagger]{Jan Kříž \orcidlink{0000-0002-8089-6729},}
\emailAdd{jan.kriz@uhk.cz}
\affiliation[\ddagger]{Department of Physics, Faculty of Science, University of Hradec Králové, \\
Rokitanského 62, 50003 Hradec Králové, Czech Republic
}

\author[*]{and Alexander Zhidenko \orcidlink{0000-0001-6838-3309}}
\emailAdd{olexandr.zhydenko@ufabc.edu.br}
\affiliation[*]{Centro de Matemática, Computação e Cognição (CMCC),\\ Universidade Federal do ABC (UFABC), \\ Rua Abolição, CEP: 09210-180, Santo André, SP, Brazil}

\arxivnumber{2501.16134}

\abstract{
Weyl gravity naturally generates effective dark matter and cosmological constant terms as integration constants, eliminating the need to explicitly introduce them into the theory. Additionally, the framework permits three intriguing solutions for compact objects: an asymptotically de Sitter Schwarzschild-like black hole described by the Mannheim-Kazanas solution, a non-Schwarzschild black hole, and a traversable wormhole that exists without exotic matter. In this work, we investigate the quasinormal spectra of all three solutions.

We demonstrate that when the mass of the black hole corresponding to the Mannheim-Kazanas solution approaches zero, the perturbation equations yield an exact solution expressible through hypergeometric functions. The quasinormal modes of black holes in Weyl gravity can be classified into three distinct branches: Schwarzschild-like modes modified by effective dark matter and cosmological terms, and modes associated with empty spacetime (de Sitter and dark matter branches), which are further influenced by the black hole mass. Previous studies have shown that the dark matter term induces a secondary stage of quasinormal ringing following the initial Schwarzschild phase. Here, we compute the frequencies using convergent methods and elucidate how this unique time-domain behavior translates into the frequency domain.

Furthermore, we demonstrate that the non-Schwarzschild black hole can be distinguished from both the Schwarzschild-like solution and the wormhole through their distinct quasinormal spectra. We also compute shadow radii for black holes and wormholes within Weyl gravity, revealing that wormholes with large throat radii can produce significantly smaller shadows compared to black holes of equivalent mass.
}

\maketitle

\section{Introduction}

The two major challenges of modern physics at large scales are explaining the accelerating separation of galaxies and the invisibility of the majority of cosmic matter within galaxies. These issues have been addressed through various, often artificial, approaches, such as introducing additional terms and parameters into Einstein's theory to replicate the observed effects.
An elegant solution to both problems was proposed long ago by Mannheim and Kazanas in \cite{Mannheim:1988dj}, where they demonstrated that pure Weyl gravity, without the Einstein term, admits a Schwarzschild-like solution incorporating effective cosmological and dark matter terms. Remarkably, these terms emerge naturally from the theory rather than being formally introduced as parameters or extra fields or distributions of matter. This is possible because the equations of motion in Weyl gravity are of fourth order, with four integration constants. Two of these constants can be associated with the dark matter and cosmological terms.

Weyl gravity, which gives rise to an effective cosmological constant and an effective dark matter term on cosmological scales, remains a viable alternative to General Relativity despite containing higher-order time derivatives. Notably, it has a well-posed initial value problem and does not introduce ghost degrees of freedom \cite{Mannheim:2000ka}. However, when considering Standard Model fields with conformal coupling, Weyl gravity predicts gravitational repulsion instead of attraction, conflicting with Solar System experiments \cite{Flanagan:2006ra,Barabash:2007ms}. Additionally, in this framework, free particles do not follow geodesic trajectories, as geodesics are not conformally invariant \cite{Mannheim:1991ez}. At the microscopic level, the theory also leads to different effective gravitational couplings for neutrons and protons \cite{Nesbet:2022rdi}, which poses further challenges for its physical viability.

In the present paper, we do not examine the issues related to conformally invariant scalar-Einstein-Weyl theories. Instead, we focus solely on classical Weyl gravity, either as a consequence of quantum conformal theory (see \cite{Mannheim:2011ds} for a review) or as an effective description of gravity at large scales. We assume that test fields are minimally coupled, ensuring that free particles follow geodesic motion. It is important to note here that the Maxwell and massless Dirac fields, when minimally coupled, retain the conformal symmetry of Weyl gravity. As a result, their behavior will differ significantly from that of a minimally coupled scalar field, which does not have the same conformal invariance.

Furthermore, as recently shown in \cite{Jizba:2024owd}, pure Weyl gravity provides a solution to a third intriguing problem: the existence of traversable wormholes without requiring exotic matter. In Einstein's gravity, traversable wormholes necessitate various forms of exotic matter, such as phantom-like fields \cite{Visser:1995cc,Blazquez-Salcedo:2018ipc,Azad:2022qqn,Battista:2024gud}, or finely tuned Maxwell-Dirac systems \cite{Konoplya:2021hsm,Blazquez-Salcedo:2020czn}. While the latter involves ordinary matter fields, it still violates traditional energy conditions. In contrast, Weyl gravity admits solutions for traversable wormholes \cite{Jizba:2024owd} as well as for non-Schwarzschild-like black holes, which differ qualitatively from the original Schwarzschild-like Mannheim-Kazanas black hole solution \cite{Mannheim:1988dj}.

One of the most important characteristics of a compact object is its quasinormal mode spectrum, which governs the evolution of perturbations around the object \cite{Nollert:1999ji, Kokkotas:1999bd, Konoplya:2011qq}. The boundary conditions for quasinormal modes of wormholes and black holes are fundamentally similar \cite{Konoplya:2005et}. Consequently, in certain parameter ranges, a wormhole may mimic a black hole \cite{Damour:2007ap}, though significant differences in their spectra \cite{Konoplya:2016hmd} and the late-time evolution of perturbations \cite{Cardoso:2016rao, Churilova:2019qph} may still exist. These distinctions make the study of perturbation evolution in Weyl gravity highly compelling.

Quasinormal modes of Schwarzschild-like Mannheim-Kazanas black holes have been investigated in a few studies \cite{Momennia:2019cfd, Momennia:2019edt, Momennia:2018hsm, Konoplya:2020fwg, Malik:2024bmp}. However, only the Schwarzschild branch of modes has been analyzed in detail. Notably, as shown in \cite{Konoplya:2020fwg}, the evolution of perturbations in this context comprises three distinct stages, each dominated by a different branch of modes: the Schwarzschild stage, the dark matter stage, and the asymptotic de Sitter stage. A comprehensive analysis of these stages in the frequency domain is still lacking. Moreover, other solutions in Weyl gravity, such as traversable wormholes and non-Schwarzschild black holes \cite{Jizba:2024owd}, have not yet been studied in terms of their quasinormal spectra.

In this work, we address these gaps by studying the time-domain evolution of perturbations and the quasinormal frequency spectra of all branches for the following compact objects in Weyl gravity:
\begin{itemize}
    \item[(a)] Mannheim-Kazanas black holes \cite{Mannheim:1988dj},
    \item[(b)] Non-Mannheim-Kazanas black holes \cite{Jizba:2024owd},
    \item[(c)] Traversable wormholes \cite{Jizba:2024owd}.
\end{itemize}

We demonstrate that for a zero-mass Mannheim-Kazanas metric, perturbation equations in the empty space of Weyl gravity allow for an exact solution, with the quasinormal modes of empty space representing the appropriate limit of compact object modes as the mass approaches zero. Additionally, we reveal several peculiarities in the spectra of these objects, which differ significantly from the spectrum of Schwarzschild spacetime in Einstein gravity. In addition, we calculate the radius of the shadow cast by all three objects and show that wormholes can be clearly distinguished from black holes in this respect.

In Section~\ref{sec:MKsolution}, we outline the fundamentals of Weyl gravity and the Mannheim-Kazanas black hole solution. Section~\ref{sec:IMsolution} provides a review of the Jizba-Mudruňka solution, which encompasses two types of objects: a black hole distinct from the Mannheim-Kazanas solution and a wormhole. Section~\ref{sec:perturbations} introduces the perturbation equations and effective potentials, while Section~\ref{sec:exactsolution} examines a special case where the mass of the Mannheim-Kazanas black hole vanishes, allowing for an exact solution. In Section~\ref{sec:numerical}, we discuss the numerical and semi-analytical methods used to determine the quasinormal frequencies. Finally, Section~\ref{sec:QNMs} analyzes the obtained quasinormal frequencies and the time-domain evolution of perturbations. Section~\ref{sec:shadows} is devoted to calculations the shadows radii for these black holes and wormholes. The conclusions (Section~\ref{sec:conclusions}) summarize our results.

\section{Weyl Gravity and Mannheim-Kazanas solution}\label{sec:MKsolution}

Within the framework of Weyl conformal gravity, the effective cosmological constant emerges naturally as an integration constant in the background solution to the vacuum field equations \cite{Mannheim:1988dj}. The corresponding action is expressed as:
\begin{align}
S = \int d^4x \sqrt{-g} \,C^{abcd}C_{abcd},
\label{Weyl-action}
\end{align}
where \( g \) denotes the determinant of the metric tensor and $C_{abcd}$ is the conformal Weyl tensor,
\begin{equation}
C_{abcd} = R_{abcd} - \frac{1}{2} \left( R_{ac} g_{bd} + R_{bd} g_{ac}- R_{ad} g_{bc}- R_{bc} g_{ad} \right) + \frac{1}{6} R \left( g_{ac} g_{bd} - g_{ad} g_{bc} \right).
 \end{equation}
The equation of motion under varying the metric is expressed via the Bach tensor,
\[
2 \partial_a \partial_d C^{ac}_{\phantom{ac}bc} + R_{ad} C^{ac}_{\phantom{ac}bc} = 0.
\]
Here \(R_{ab}\) is the Ricci tensor.
Notably, Birkhoff's theorem holds true within Weyl conformal gravity \cite{Riegert:1984zz}.
A static and spherically symmetric vacuum solution representing a black hole in this theory was first derived in \cite{Mannheim:1988dj}.

The Mannheim-Kazanas solution is a solution to the field equations derived from the conformal gravity action, which fundamentally differ from Einstein's equations. This solution was introduced by P. Mannheim and D. Kazanas in the context of efforts to explain galactic rotation curves without invoking dark matter. Unlike standard General Relativity, which relies on the Einstein-Hilbert action, conformal gravity is based on the Weyl action. The Weyl action is invariant under conformal transformations, meaning it remains unchanged under local rescaling of the metric \( g_{\mu\nu}(x) \to \Omega^2(x) g_{\mu\nu}(x) \), where \( \Omega(x) \) is a smooth, nonzero function of spacetime coordinates.

The Mannheim-Kazanas solution is a static, spherically symmetric metric derived from conformal gravity. The spacetime interval in this solution is \cite{Mannheim:1988dj}:
\begin{equation}
ds^2 = -f(r) dt^2 + \frac{dr^2}{f(r)} + R(r)^2 \left( d\theta^2 + \sin^2\theta \, d\phi^2 \right),
\end{equation}
where the metric functions are given by:
\begin{equation}\label{MKsolution}
f(r) = 1 - \alpha - \frac{2 M}{r} + \gamma r - k r^2, \qquad R(r)=r.
\end{equation}
Since Mannheim and Kazanas considered the neutral black holes, the values of $\alpha$ and $M$ are not independent and can be defined through the new parameter $\beta$, as follows:
\begin{eqnarray}\label{MKrelationa}
    \alpha&=&3\beta\gamma,
    \\\label{MKrelationM}
    M&=&\beta\left(1-\frac{3\beta\gamma}{2}\right).
\end{eqnarray}
When $\gamma=0$, $M=\beta$ and the Mannheim-Kazanas black hole coincides with the Schwarzschild-de Sitter solution in General Relativity.

The $\gamma r$ term is particularly significant in explaining the flat rotation curves of galaxies. In standard GR, flat rotation curves require additional mass (dark matter) at large distances from the galactic center. In conformal gravity, the \( \gamma r \) term provides a natural explanation for the observed rotation curves without requiring dark matter.

The \( -k r^2 \) term is associated with a cosmological contribution and is negligible on galactic scales. It becomes relevant at even larger scales, possibly connecting to cosmological acceleration. The Mannheim-Kazanas solution has been used to fit galactic rotation curves with only ordinary (baryonic) matter. Although the solution can fit many galactic rotation curves, its universal applicability to all galaxies and consistency with other gravitational phenomena (e.g., lensing, galaxy clusters) requires further scrutiny.

Various properties and generalizations of the Mannheim-Kazanas black holes, with an emphasis on particle motion, gravitational lensing, and Hawking radiation, have been studied in \cite{Islam:2018ymd,Dutta:2018oaj,Takizawa:2020dja,Kasikci:2018mtg,Fathi:2020sey,Fathi:2019jgd,Fathi:2020sfw,Xu:2018liy,Momennia:2019cfd}.

\section{Jizba-Mudruňka solution}\label{sec:IMsolution}

Recently, another class of spherically symmetric solutions in the Weyl gravity was obtained using Newman-Penrose formalism \cite{Jizba:2024owd}. The metric function of these solutions has the form:
\begin{equation}\label{JMsolution}
f(r) = R^2(r)\left(\frac{1 - \alpha + kr_0^2}{\rho^2(r)} - \frac{6 M+2\gamma r_0^2}{3\rho^3(r)} + \frac{\gamma}{\rho(r)} - k\right), \qquad R(r)=\sqrt{r^2+r_0^2},
\end{equation}
where one fixes
\begin{equation}\label{rhodef}
\frac{1}{\rho(r)}=\intop_r^{\infty}\frac{dr}{R^2(r)}=\frac{1}{r_0}\left(\frac{\pi}{2}-\arctan{\frac{r}{r_0}}\right),
\end{equation}
so that, for $r\gg r_0$, we have
\[
\frac{\rho(r)}{r}=1+\Order{\frac{r_0}{r}}^2.
\]

Therefore, asymptotically, we have
\begin{equation}
f(r)=1 - \alpha - \frac{2 M}{r} + \gamma r - k r^2 + \Order{\frac{r_0}{r}}^2,
\end{equation}
and one recovers the Mannheim-Kazanas solution as $r_0\to0$.

On the other hand, we have
\begin{equation}
    \lim_{r\to-\infty}\rho(r)=\frac{r_0}{\pi},
\end{equation}
and
\begin{equation}\label{-infinity}
\lim_{r\to-\infty}\frac{f(r)}{r^2}=\pi^2\frac{1 - \alpha}{r_0^2}-\frac{\pi\gamma}{r_0}\left(\frac{2\pi^2}{3}-1\right)- \frac{2 M\pi^3}{r_0^3} + k\left(\pi^2-1\right).
\end{equation}

The electric charge of the Maxwell field $Q$,
\(
{\cal A}_adx^a=\frac{Q}{R(r)}dt,
\)
satisfies \cite{Jizba:2024owd}
\begin{equation}
    1-\left(1-\alpha+kr_0^2\right)^2-6M\gamma-2\gamma^2r_0^2=\frac{3G_W^2Q^2}{2}.
\end{equation}
Therefore, if we consider the uncharged solution, for consistency with eq.~(\ref{MKsolution}), we will define $M$ and $\alpha$ through the parameter $\beta$, as follows:
\begin{equation}\label{betadef}
\begin{array}{rcl}
    \alpha&=&3\beta\gamma + k r_0^2,
    \\
    M&=&\beta\left(1-\dfrac{3\beta\gamma}{2}\right) -\dfrac{\gamma r_0^2}{3}.
\end{array}
\end{equation}
Then, in the limit $r_{0}\to0$, relations (\ref{MKrelationa}) and (\ref{MKrelationM}) are reproduced.

The parameter $r_0$ is the minimum value of $R(r)$, which corresponds to $r=0$. Requiring that the metric function $f(r)$ must be positive for a wormhole solution at $r=0$, we see that for $M/r_{0}$ smaller than,
\begin{equation}\label{inequality1}
\frac{M}{r_0}<\frac{1 - \alpha}{\pi}-\gamma r_0\left(\frac{1}{3}-\frac{2}{\pi^2}\right)+kr_0^2\frac{\pi^2-4}{\pi^3},
\end{equation}
$r=0$ is the position of the wormhole throat.

At $r\to- \infty$ the asymptotic of the wormholes obeying the above inequality (\ref{inequality1} ) may be either de Sitter or anti-de Sitter one, depending on the sign of $f(r)/r^2$ in this limit. If, in addition to eq.~(\ref{inequality1}), we have
\begin{equation}
\frac{M}{r_0} > \frac{1 - \alpha}{2\pi}-\gamma r_0\left(\frac{1}{3}-\frac{1}{2\pi^2}\right)+kr_0^2\frac{\pi^2-1}{2\pi^3},
\end{equation}
then, the wormhole's throat is a passage to the de Sitter universe.
When, otherwise, the following inequality is satisfied,
\begin{equation}
\frac{M}{r_0}<\frac{1 - \alpha}{2\pi}-\gamma r_0\left(\frac{1}{3}-\frac{1}{2\pi^2}\right)+kr_0^2\frac{\pi^2-1}{2\pi^3},
\end{equation}
then, the asymptotic values of $f(r)/r^2$ is larger than zero, and we have the anti-de Sitter (AdS) asymptotic.

A symmetric wormhole is possible only with two AdS asymptotics on both sides of the throat for a specific relation between the constants $M$ and $\gamma$,
\begin{equation}\label{necessary}
M=-\frac{\gamma r_0^2}{3}.
\end{equation}
Indeed, then the second term in the metric function (\ref{JMsolution}) vanishes, which is the necessary, but not sufficient, condition for the symmetry of the metric relatively the throat.

For the neutral case, there is a symmetric wormhole with a negative asymptotic mass, which was considered in \cite{Jizba:2024owd}, provided
\[
\gamma=\frac{2}{3\beta}=\frac{\pi}{r_0} \quad\Longrightarrow\quad M=-\frac{\pi}{3}r_0<0,
\]
then eq.~(\ref{necessary}) is satisfied and
\begin{eqnarray}
    f(r)=\frac{r^2+r_0^2}{L^2}+\frac{r^2+r_0^2}{r_0^2}\left(\frac{\pi^2}{4}-\arctan^2{\frac{r}{r_0}}\right)
    \\\nonumber\left(k\equiv-\frac{1}{L^2},\quad\alpha=2-kr_0^2=2+\frac{r_0^2}{L^2}\right).
\end{eqnarray}

There is also a symmetric wormhole solution with positive mass for
\[
\beta=0, \quad M=-\frac{\gamma r_0^2}{3}=\frac{\pi r_0}{3}>0.
\]
In this case, we have
\begin{eqnarray}
    f(r)=\frac{r^2+r_0^2}{L^2}-\frac{r^2+r_0^2}{r_0^2}\left(\frac{\pi^2}{4}-\arctan^2{\frac{r}{r_0}}\right)
    \\\nonumber\left(k\equiv-\frac{1}{L^2},\quad\alpha=-kr_0^2=\frac{r_0^2}{L^2}\right).
\end{eqnarray}
Here we have a wormhole only when
\[
r_0>\frac{\pi L}{2}.
\]
For smaller values of $r_0$, the throat is hidden by the event horizons, located at $r=\pm r_h$, where
\[
r_h=r_0\tan\sqrt{\frac{\pi^2}{4}-\frac{r_0^2}{L^2}},
\]
and we obtain two asymptotically AdS black holes.
It is interesting to note that for $r_0\ll L$, we have
\[
r_h=\frac{L^2}{r_0}+\Order{r_0},
\]
so that the horizons approach the AdS bounds, when decreasing $r_0$.

In the present work, we focus on perturbations around asymptotically flat or de Sitter black holes and wormholes, leaving the more exotic anti-de Sitter case for future investigation.
While the expected values of the cosmological constant (related to $k$) and the term explaining the rotation curves of galaxies $\gamma$ should be relatively small (see, for example \cite{Diaferio:2008gh,Horne:2016ajh,Yang:2013skk}) to produce pronounced effect in the observed quasinormal spectrum, we study the perturbations for all allowed values of these parameters, implying not only theoretically full description of the spectra, but also meaning that the ratio between the constants could be different in the early Universe.  

\section{Wavelike equations and effective potentials}\label{sec:perturbations}

Gravitational perturbations described by a fourth-order differential equation present a highly challenging problem, raising fundamental questions about the nature of perturbations’ propagation. In particular, within the conformal Einstein-scalar-Weyl theory, gravitational perturbations include not only the standard transverse massless modes but also additional massive propagating modes \cite{Caprini:2018oqe}. These massive modes can alter the behavior of gravitational waves at different energy scales, potentially leading to observable deviations from General Relativity in astrophysical and cosmological settings \cite{Holscher:2019swu}.

As a first step toward a comprehensive analysis of spacetime perturbations, we will focus on perturbations of test scalar, electromagnetic, and neutrino fields. In many cases, the evolution of gravitational field perturbations is qualitatively similar to that of matter fields. Indeed, in the high-frequency regime ($\ell \rightarrow \infty$), quasinormal modes typically do not depend on the spin of the field. However, exceptions have been observed, as described in \cite{Konoplya:2017wot}, and are often associated with higher-order curvature terms \cite{Konoplya:2020bxa,Bolokhov:2023dxq} or cosmological factors \cite{Konoplya:2022gjp}. Therefore, we cannot rule out the possibility that gravitational perturbations introduce unique features into the quasinormal spectrum.

The scalar ($\Phi$), electromagnetic ($A_\mu$), and Dirac ($\Psi_{Dirac}$) fields in a curved spacetime obey the general covariant equations:
\begin{subequations}\label{coveqs}
\begin{eqnarray}\label{KGg}
\frac{1}{\sqrt{-g}}\partial_\mu \left(\sqrt{-g}g^{\mu \nu}\partial_\nu\Phi\right)&=&0,
\\\label{EmagEq}
\frac{1}{\sqrt{-g}}\partial_{\mu} \left(F_{\rho\sigma}g^{\rho \nu}g^{\sigma \mu}\sqrt{-g}\right)&=&0\,,
\\\label{covdirac}
\gamma^{\alpha} \left( \frac{\partial}{\partial x^{\alpha}} - \Gamma_{\alpha} \right) \Psi_{Dirac}&=&0,
\end{eqnarray}
\end{subequations}
where $F_{\mu\nu}=\partial_\mu A_\nu-\partial_\nu A_\mu$ is the electromagnetic tensor, $\gamma^{\alpha}$ are (noncommutative) gamma matrices and $\Gamma_{\alpha}$ are spin connections.
After separation of variables and implying static, spherically symmetric background, these equations (\ref{coveqs}) can be reduced to the wavelike form with an effective potential \cite{Kokkotas:1999bd,Berti:2009kk,Konoplya:2011qq}:
\begin{equation}\label{wave-equation}
\dfrac{d^2 \Psi}{dr_*^2}+(\omega^2-V(r))\Psi=0,
\end{equation}
where the ``tortoise coordinate'' $r_*$ is defined as follows
\begin{equation}\label{tortoise}
dr_*\equiv\frac{dr}{f(r)}.
\end{equation}

From (\ref{JMsolution}) it follows that the expression
\begin{equation}\label{Pdef}
\frac{f(r)}{R^2(r)} = P(\rho(r))
\end{equation}
depends on $\rho(r)$ only. Here we have introduced the function
\begin{equation}\label{Prho}
    P(\rho) \equiv \frac{1 - \alpha + kr_0^2}{\rho^2} - \frac{6 M+2\gamma r_0^2}{3\rho^3} + \frac{\gamma}{\rho} - k = \frac{1 - 3\beta\gamma}{\rho^2} - \frac{2\beta-3\beta^2\gamma}{\rho^3} + \frac{\gamma}{\rho} - k,
\end{equation}
which does not depend on $r_0$ once $\alpha$ and $M$ are given in terms of the parameter $\beta$ through (\ref{betadef}).

Therefore, the tortoise coordinate is also a function of $\rho$,
\begin{equation}\label{tortoiseP}
dr_*=\frac{dr}{f(r)}=\frac{dr}{R^2(r)P(\rho(r))}=\frac{d\rho}{\rho^2P(\rho)}.
\end{equation}

The effective potentials for scalar ($s=0$) and electromagnetic ($s=1$) fields can be written in the following form,
\begin{equation}\label{potentialScalar}
V(r)=f(r)\frac{\ell(\ell+1)}{R(r)^2}+\frac{1-s}{R(r)}\cdot\frac{d^2 R}{dr_*^2} = \ell(\ell+1)P(\rho(r))+\frac{1-s}{R(r)}\cdot\frac{d^2 R}{dr_*^2},
\end{equation}
where $\ell=s, s+1, s+2, \ldots$ is the multipole number appearing as a result of separation of angular variables, and $s=0$ ($s=1$) for the scalar (electromagnetic) field respectively.

It is interesting to note that the effective potential for $s=1$ is proportional to $P(\rho)$, and, since the tortoise coordinate is also defined in (\ref{tortoiseP}) through $P(\rho)$, the wavelike equation (\ref{wave-equation}) does not depend on $r_0$ when we express $M$ in terms of the parameter $\beta$.

For the Dirac field we have two isospectral potentials, corresponding to the two degrees of freedom (chiralities),
\begin{equation}
V_{\pm}(r) = W^2\pm\frac{dW}{dr_*}, \quad W\equiv \left(\ell+\frac{1}{2}\right)\frac{\sqrt{f(r)}}{R(r)}=\left(\ell+\frac{1}{2}\right)\sqrt{P(\rho(r))}.
\end{equation}
The corresponding isospectral solutions of the Dirac equation are linked by the Darboux transformation,
\begin{equation}\label{psi}
\Psi_{+}\propto \left(W+\dfrac{d}{dr_*}\right) \Psi_{-}.
\end{equation}
Consequently, from the point of view of quasinormal mode calculations, it is sufficient to use only one of the effective potentials. We use here $V_{+}(r)$, because the WKB method is usually more accurate for such choice. Since both potentials are again defined through $P(\rho)$ and its derivatives with respect to the tortoise coordinate, we conclude that the perturbation equation for the test Dirac field does not depend on $r_0$ when we parametrize $M$ via the parameter $\beta$.

Thus, only the minimally coupled test scalar field depends on the wormhole's throat size $r_0$ in the chosen parametrization of the metric.

If the effective potentials is positive definite, the corresponding differential operator,
\[
-\frac{d^2}{d r_*^2} + V(r_*)
\]
is a positive self-adjoint operator in the Hilbert space of square integrable functions of the tortoise coordinate  \(r_*\). As a result, all solutions of the linearized dynamical field equations with compact support initial conditions are bounded. In other words, the wave function is bounded in time and the perturbation is stable. Figures \ref{fig:flats0l0} and \ref{fig:flats0l1} show that, for certain parameter values, the effective potential for the scalar field can exhibit a negative gap. This indicates that the stability of the perturbations is not guaranteed, necessitating time-domain integration of the wave equations to observe the behavior up to the stage of asymptotic tails.

It is important to note that, although the perturbation equations for the Maxwell and Dirac fields in terms of the parameter $\beta$ do not depend on $r_0$, being essentially the same for the Mannheim-Kazanas black holes and Jizba-Mudruňka black holes and wormholes, the asymptotic mass depends on the value of the throat size $r_0$, unless $\gamma=0$. If the dark matter term $\gamma\neq0$, the quasinormal modes measured in units of the asymptotic mass, $\omega M$, depend on the value of $r_0$. Namely, both the oscillation frequency and the damping rate increase with $r_0$ when $\gamma<0$ and decrease for $\gamma>0$:
\[
\omega M = \omega \beta \left(1-\frac{3\beta\gamma}{2} -\frac{\gamma r_0^2}{3\beta}\right).
\]

\section{Exact solutions in the limit of vanishing black-hole mass}\label{sec:exactsolution}

Here we will consider a few cases of vacuum spacetimes allowing for exact analytic solutions. We will be limited by those values of the parameters $\gamma$ which provide asymptotically de Sitter-like solution with a null hypersurface in the far zone, which is similar to the cosmological horizon, rather than anti-de Sitter like asymptotic. Therefore, for example, when the cosmological constant vanishes ($k=0$), we are limited by negative values of $\gamma$.

\subsection{Zero cosmological constant, $\gamma<0$}

In the limit of vanishing effective cosmological constant $k=0$ and black-hole mass $M=0$ ($\beta=0$, $\alpha=0$) for $\gamma<0$, the Mannheim-Kazanas (\ref{MKsolution}) metric function takes the simple form:
\begin{equation}\label{MKzero}
f(r) = 1 - \frac{r}{r_c},
\end{equation}
where $r_c=-1/\gamma$ is the cosmological horizon. Note, that this horizon appears without the effective lambda-term ($k=0$) due to the negative value of $\gamma$, which governs the accelerated expansion of the universe.

The derivative of the metric function (\ref{MKzero}) is a constant, and the perturbation equations are greatly simplified
\begin{equation}\label{simpleeq}
\left(1 - \frac{r}{r_c}\right) \left(\left(1 - \frac{r}{r_c}\right) \Psi ''(r)-\frac{1}{r_c} \Psi'(r)\right)+\left(w^2-\left(1 - \frac{r}{r_c}\right)\frac{\ell(\ell+1)-(1-s)r/r_c }{r^2}\right)\Psi (r)=0
\end{equation}
and its general solution for the electromagnetic perturbations ($s=1$) is
\begin{eqnarray}
    \Psi(r)=\left(1 - \frac{r}{r_c}\right)^{-\imo\omega r_c}\Biggl(&&C_1\cdot r^{-\ell}\operatorname{_2F_1}\left(-\ell,-\ell-2\imo\omega r_c,-2\ell ~|~ r/r_c\right)
\\\nonumber&+&C_2\cdot r^{\ell+1}\operatorname{_2F_1}\left(\ell+1,\ell+1-2\imo\omega r_c,2\ell+2~|~ r/r_c\right)\Biggr),
\end{eqnarray}
where $\operatorname{_2F_1}(a,b,c~|~z)$ is the hypergeometric function, which is regular at $r=0$. Therefore, for the regular solutions we take $C_1=0$.

Since $\ell+1>0$ the quasinormal boundary conditions at $r\to r_c-0$ (purely outgoing wave) are satisfied
iff $\ell+1-2\imo\omega r_c=-n$, where $n$ is nonnegative integer. Thus, we obtain the quasinormal spectrum,
\begin{equation}\label{MKzeroEM}
    \omega_n=-\imo\frac{\ell+1+n}{2r_c}, \quad n=0,1,2,\ldots.
\end{equation}

For the scalar field ($s=0$), the general solution can be represented in a similar form,
\begin{eqnarray}
    \Psi(r)=\left(1 - \frac{r}{r_c}\right)^{-\imo\omega r_c}\Biggl(&&C_1\cdot r^{-\ell}\operatorname{_2F_1}\left(-\ell+\frac{1}{2\imo\Omega r_c},-\ell-2\imo\Omega r_c,-2\ell ~|~ r/r_c\right)
\\\nonumber&+&C_2\cdot r^{\ell+1}\operatorname{_2F_1}\left(\ell+1+\frac{1}{2\imo\Omega r_c},\ell+1-2\imo\Omega r_c,2\ell+2~|~ r/r_c\right)\Biggr),
\end{eqnarray}
where
\[
\Omega=\omega\frac{1+\sqrt{1-\omega^{-2}r_c^{-2}}}{2},
\]
and the quasinormal modes can be given by the following expression:
\begin{equation}\label{MKzeroS}
    \omega_n=-\frac{\imo}{2r_c}\left(\ell+1+n-\frac{1}{\ell+1+n}\right), \quad n=0,1,2,\ldots.
\end{equation}
One should note that, for $\ell=n=0$, formula~(\ref{MKzeroS}) gives the eigenvalue $\omega=0$, for which $\Omega=-\imo/2r_c$. The corresponding algebraically special solution $\Psi(r)=C_2\cdot r$ is not dynamic, being a constant contribution to the scalar field in (\ref{KGg}), $\Phi\propto\Psi/r$, which redefines the vacuum state of the scalar field.

Once we exclude $\ell=n=0$ from the quasinormal spectrum, the fundamental mode of the $\ell=0$ spectrum corresponds to $n=1$ in~(\ref{MKzeroS}) and we can rewrite the expression ~(\ref{MKzeroS}) for $\ell=0$ as follows
\begin{equation}\label{MKzeroSl0}
    \omega_{n,\ell=0}=-\frac{\imo}{2r_c}\left(2+n-\frac{1}{2+n}\right), \quad n=0, 1,2,\ldots.
\end{equation}
which evidently coincides with expression~(\ref{MKzeroS}) for $\ell=1$.
Therefore, quasinormal modes of the test scalar field for $\ell=0$ and $\ell=1$ coincide in the limit of vanishing black-hole mass.

\subsection{Nonzero cosmological constant}

When $k$ is nonzero, we define
\begin{equation}
\gamma=-\frac{1}{r_c}+k r_c,
\end{equation}
and assume that $k>-r_c^{-2}$, so that $r_c$ is again the cosmological horizon,
\begin{equation}\label{MKk}
f(r) = \left(1 - \frac{r}{r_c}\right)(1+k r_c \cdot r).
\end{equation}
It is interesting to note that after introducing the new variable
\begin{equation}
\widetilde{r}=r\cdot\frac{1+kr_c^2}{1+k r_c \cdot r}, \qquad 0\leq \widetilde{r}\leq r_c,
\end{equation}
the wavelike equation for $s=1$ takes the form of (\ref{simpleeq}) with the effective frequency
\begin{equation}
\widetilde{\omega}=\frac{\omega}{1+kr_c^2}.
\end{equation}
Therefore, we conclude that, for the electromagnetic perturbations,
\begin{equation}\label{MKspectrumEM}
    \omega_n=(1+kr_c^2)\widetilde{\omega}=-\imo\frac{1+kr_c^2}{2r_c}\left(\ell+1+n\right).
\end{equation}

When $k=r_c^{-2}$ then $\gamma=0$, and (\ref{MKspectrumEM}) is reduced to the de Sitter spectrum \cite{Lopez-Ortega:2012xvr,Lopez-Ortega:2007vlo}:
\begin{equation}
\omega_n = -\imo\frac{\ell+1+n}{r_c} \quad (s=1).
\end{equation}
Notice that the de Sitter spectrum for $\gamma=0$ differs from the one for the de Sitter-like universe ($k=0$, $\gamma<0$), given in geometric units of the horizon radius $r_c$ by Eq.~(\ref{MKzeroEM}).

For $s=0$ the equation has an additional singular point and the solution cannot be expressed in terms of the hypergeometric function. Analysis of the corresponding Heun equation is beyond the scope of the present paper.

In sec.~\ref{sec:QNMs}, we will show that for those cases when we were able to find the exact analytic solutions of the vacuum spacetime, quasinormal modes of black holes tend to those of the vacuum when mass of the black hole goes to zero. This could be expected by analogy with \cite{Konoplya:2022kld} where it was found that for general spherically symmetric and asymptotically de Sitter black holes in metric theories of gravity we have
\begin{equation}
\omega_n = \omega_n^{(dS)} \left( 1 - \frac{r_0 (1 + \epsilon)}{2 r_c} + \mathcal{O}\left(\frac{r_e}{r_c}\right)^2 \right)
= \omega_n^{(dS)} \left( 1 - \frac{M}{r_c} + \mathcal{O}\left(\frac{M}{r_c}\right)^2 \right),
\end{equation}
where $\epsilon$ is deviation of the event horizon radius from its Schwarzschild limit. This relation is valid when the dark matter term vanishes, but when $\gamma\neq0$, the correction terms containing the black hole mass must be different.

\section{Shadows}\label{sec:shadows}

\begin{figure}
\resizebox{\linewidth}{!}{\includegraphics{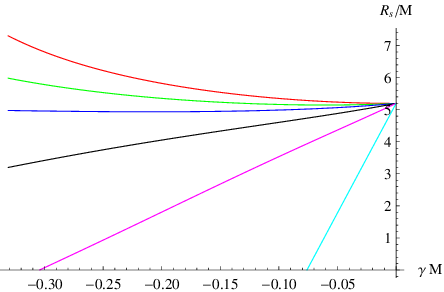}}
\caption{Radius of the shadow cast by a black hole or wormhole as a function of $\gamma$ for various values of $r_0$: $r_0=0$ (red, top), $r_0=M$ (green), $r_0=1.5M$ (blue), $r_0=2M$ (black), $r_0=3.14M$ (pink), $6.28M$ (bottom, cyan); $k=0$.}\label{fig:Shadows}
\end{figure}

Shadows cast by black holes in the presence of dark matter have been extensively studied in numerous publications, making it difficult to review the entire body of work within the scope of this paper. Here, we highlight a selection of studies that employ various approaches to modeling dark matter \cite{Konoplya:2019sns,Jusufi:2019nrn,Hou:2018avu,Nampalliwar:2021tyz,Cardoso:2021wlq,Rahaman:2021web,Kasuya:2021cpk,Capozziello:2023tbo,Figueiredo:2023gas,Chen:2024lpd,Pantig:2024rmr,Gomez:2024ack,Macedo:2024qky}, as well as several works that develop general formalisms for determining the parameters of black hole and wormhole shadows \cite{Perlick:2021aok,Younsi:2016azx,Bronnikov:2021liv,Konoplya:2021slg}.

The literature on wormhole shadows is similarly extensive (see, for example, \cite{Nedkova:2013msa,Shaikh:2018kfv,Ohgami:2015nra,Amir:2018pcu,Churilova:2021tgn} and references therein). A comprehensive analysis of various wormhole configurations, as presented in \cite{Bronnikov:2021liv}, could also prove valuable for further research.

We consider the motion in the equatorial plane ($\theta=\pi/2$, $d\theta=0$) along the null geodesic,
\begin{eqnarray}\label{nullgeodesic}
ds^2&=&-f(r)dt^2+\frac{dr^2}{f(r)}+R(r)^2d\phi^2=0,\\
\frac{d^2r}{dt^2}&=&-\frac{f(r)f'(r)}{2}dt^2+\frac{f'(r)}{2f(r)}dr^2+f(r)R(r)R'(r)d\phi^2.
\end{eqnarray}
The circular orbit $r=r_c$, $dr=0$, $d^2r=0$, satisfies
\begin{equation}\label{photonorbit}
    f'(r_c)=2f(r_c)\frac{R'(r_c)}{R(r_c)},
\end{equation}
and the corresponding angular frequency,
\begin{equation}\label{photonorbitOmega}
\Omega_c=\frac{d\phi}{dt}=\frac{\sqrt{f(r_c)}}{R(r_c)}.
\end{equation}
We note that (\ref{photonorbit}) is the extremum condition of the righthand side of (\ref{photonorbitOmega}), therefore,
\begin{equation}\label{photonorbitmax}
\Omega_c=\max_r\frac{\sqrt{f(r)}}{R(r)}=\max_{\rho}\sqrt{P(\rho)},
\end{equation}
where the function $P(\rho)$ given by (\ref{Prho}) in terms of the parameter $\beta$ does not depend on $r_0$
\begin{equation}
    P(\rho) = \frac{1 - 3\beta\gamma}{\rho^2} - \frac{2\beta-3\beta^2\gamma}{\rho^3} + \frac{\gamma}{\rho} - k.
\end{equation}

The maximum of the function $P(\rho)$ is reached at $\rho=3\beta$. Finally, we find
\begin{equation}\label{Omegac}
\Omega_c=\sqrt{P(3\beta)}=\sqrt{\frac{1+3\beta\gamma}{27\beta^2}-k}.
\end{equation}

The shadow radius visible by a remote observer,
\begin{equation}\label{shadowdef}
R_s\equiv\frac{1}{\Omega_c}=\frac{1}{\sqrt{P(3\beta)}}=\frac{3\sqrt{3}\beta}{\sqrt{1+3\beta\gamma-27k\beta^2}}.
\end{equation}

When $\gamma=k=0$ ($\beta=M$) we reproduce the well known formula for the radius of the shadow of the Schwarzschild black hole. In fig. \ref{fig:Shadows} the radius of the shadow is shown in units of the asymptotic black hole/wormhole mass $M$. There we can see that once $r_0$ is small, the radius of the shadow grows once the absolute value of $\gamma$ is increased. However, at some $r_0$ the situation changes to the opposite: larger absolute values of $\gamma$ lead to smaller radius of the shadow, so that for the near extreme wormholes with $r_0= 6.28M$, the radius of the shadow goes to zero at some critical values of $\gamma$ corresponding to the limit $\beta\to0$.

By substituting $r=r_c+\delta r$ into (\ref{nullgeodesic}), we obtain the equation for the radial coordinate of the photons that are leaving the circular orbit,
\begin{equation}
\left(\frac{d}{dt}\delta r\right)^2=\lambda^2\delta r^2+\Order{\delta r}^3,
\end{equation}
where $\lambda$ is the the Lyapunov exponent, satisfying,
\begin{equation}\label{Lyapunovdef}
\lambda^2=-\frac{1}{2P(\rho(r))}\frac{d^2P}{dr_*^2}\Biggr|_{r=r_c}=-\frac{\rho^4P(\rho)P''(\rho)}{2}\Biggr|_{\rho=3\beta}=\frac{1+3\beta\gamma}{27\beta^2}-k.
\end{equation}

\section{Numerical methods for the calculation of quasinormal modes}\label{sec:numerical}

The quasinormal modes of asymptotically flat black holes are the eigenvalues of the wavelike equation (\ref{wave-equation}) with the boundary conditions:
\begin{equation}\label{boundaryconditions}
\Psi(r_*\to\pm\infty) \propto e^{\pm \imo \omega r_*},
\end{equation}
which correspond to a purely ingoing wave at the event horizon ($r_*\to-\infty$) and a purely outgoing wave at spatial infinity ($r_*\to\infty$).
When the spacetime is asymptotically de Sitter one, the outgoing boundary condition is imposed on the de Sitter horizon instead.

The same boundary conditions are appropriate to traversable wormholes, where for asymptotically flat spacetimes, $r_*\to-\infty$ corresponds to spatial infinity in the other universe \cite{Konoplya:2005et}.
If a wormhole is asymptotically de Sitter at one or both asymptotics, then, again, we imply purely outgoing boundary conditions at the de Sitter horizon. The similarity in boundary conditions for black holes and wormholes allows one to use the same techniques for finding quasinormal of both compact objects.

The boundary conditions (\ref{boundaryconditions}) imply that the time dependence of the perturbations is given by $\Psi\propto e^{-\imo\omega t}$, so that the modes with $\im{\omega} < 0$ correspond to decaying signal. Due to the symmetry of Eq.~(\ref{wave-equation}), all solutions to the eigenvalue problem with a negative real part of the frequency can be obtained by taking the complex conjugate of the corresponding solutions with a positive real part. Consequently, in our analysis, we assume $\re{\omega} > 0$ without loss of generality.

We provide a brief overview of the WKB approximation, time-domain integration, and the Bernstein polynomial methods, which are employed to compute the quasinormal frequencies of black holes and wormholes.

\subsection{WKB approach}\label{sec:WKB}

\begin{figure*}
\resizebox{\linewidth}{!}{\includegraphics{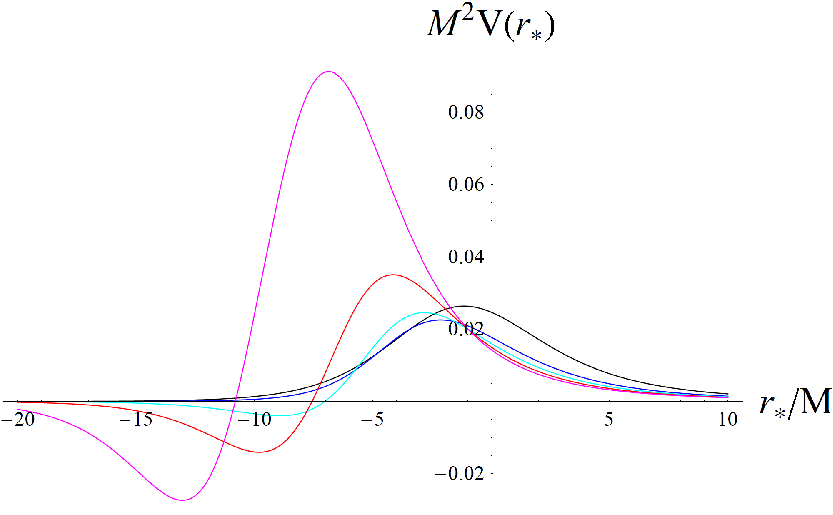}\includegraphics{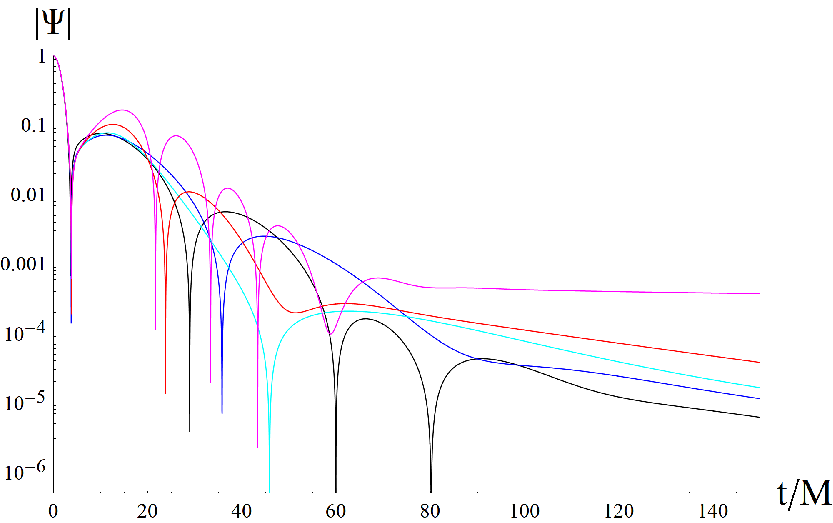}\includegraphics{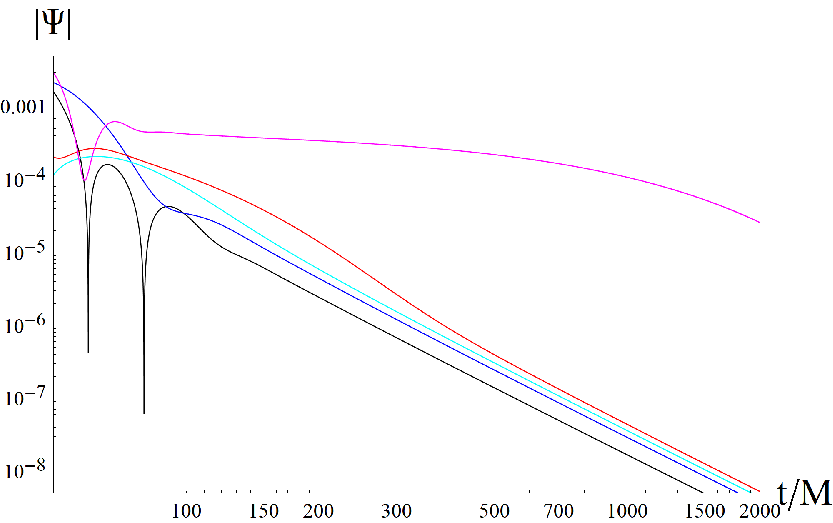}}
\caption{The effective potentials (left panel) and time-domain profiles (semi-log plot for the ringdown and log-log plot for the late-time tails) for the scalar perturbations ($\ell=0$) of the asymptotically flat Jizba-Mudruňka solution ($k=0$, $\gamma=0$, $\alpha=0$, $M=\beta$): $r_0=0$ (black), $r_0=3M$ (blue), $r_0=4M$ (cyan), $r_0=5M$ (red), $r_0=6M$ (magenta). $r_0=0$ corresponds to the Schwarzschild black hole, for $0<r_0\leq\pi M$ there is a black hole, which differs from Schwarzschild, when $\pi M<r_0<2\pi M$ there is a throat, which connects the flat space to a de Sitter universe. The time-domain profile is calculated for $r=3M$ ($r_*=0$).}\label{fig:flats0l0}
\end{figure*}

\begin{figure*}
\resizebox{\linewidth}{!}{\includegraphics{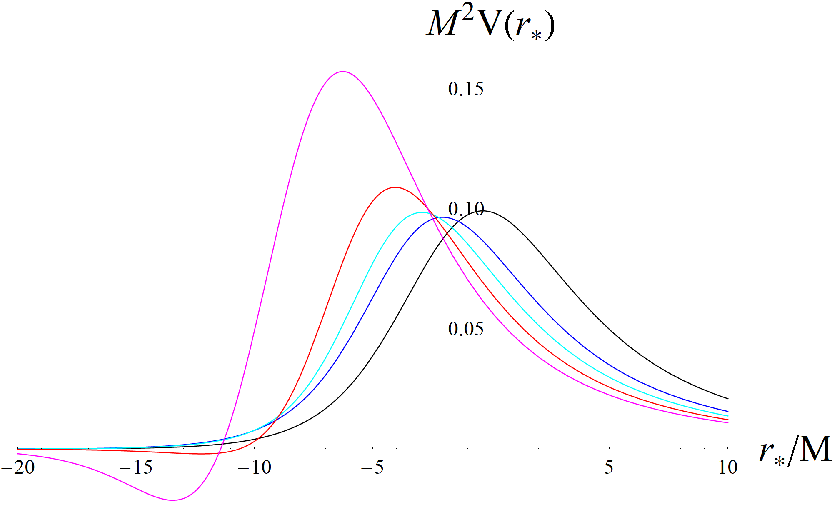}\includegraphics{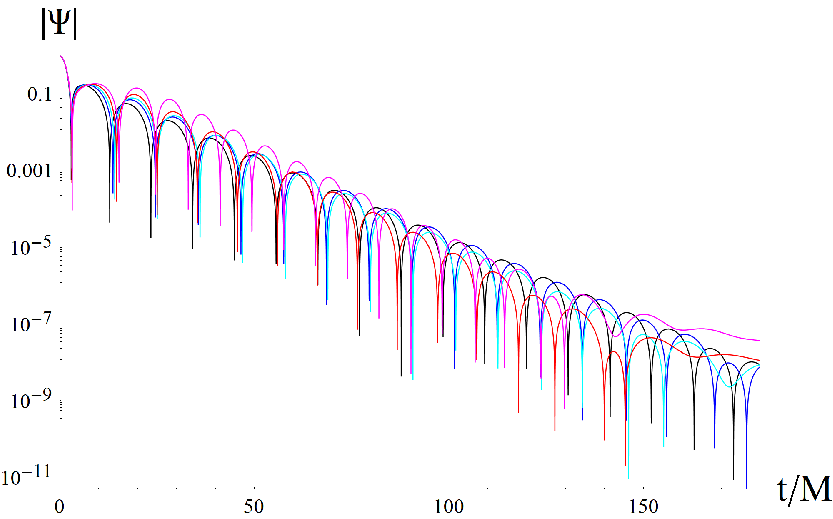}\includegraphics{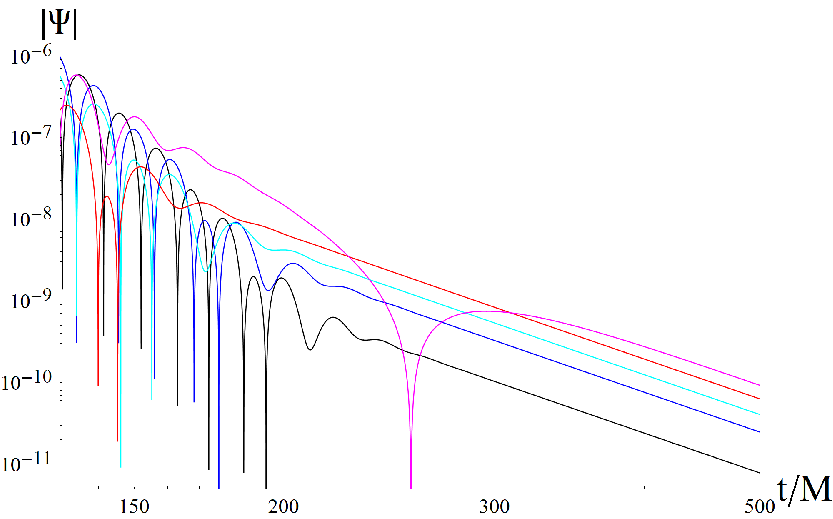}}
\caption{The effective potentials (left panel) and time-domain profiles (semi-log plot for the ringdown and log-log plot for the late-time tails) for the scalar perturbations ($\ell=1$) of the asymptotically flat Jizba-Mudruňka solution ($k=0$, $\gamma=0$, $\alpha=0$, $M=\beta$): $r_0=0$ (black), $r_0=3M$ (blue), $r_0=4M$ (cyan), $r_0=5M$ (red), $r_0=6M$ (magenta). $r_0=0$ corresponds to the Schwarzschild black hole, for $0<r_0\leq\pi M$ there is a black hole, which differs from Schwarzschild, when $\pi M<r_0<2\pi M$ there is a throat, which connects the flat space to a de Sitter universe. The time-domain profile is given for $r=3M$ ($r_*=0$).}\label{fig:flats0l1}
\end{figure*}

The WKB method is one of the most widely used techniques for calculating low-lying quasinormal modes, owing to its speed, automation, and generally sufficient accuracy in most cases (see \cite{Barrau:2019swg,Stashko:2024wuq,Dubinsky:2024hmn,Zinhailo:2019rwd,Xiong:2021cth,Skvortsova:2024atk,Skvortsova:2024wly,Malik:2024nhy,Hamil:2024nrv,Liu:2024wch} for recent examples of applications). This method is based on matching the asymptotic solutions, which satisfy the quasinormal boundary conditions (\ref{boundaryconditions}), with the Taylor expansion of the potential around its peak, through the two turning points defined by the equation
\[
V(r_*) = \omega^2.
\]
As a result, the WKB method is particularly effective when the effective potential has a barrier shape with a single peak. Even in cases where a small negative gap exists near the event horizon, as illustrated in Figures \ref{fig:flats0l0} and \ref{fig:flats0l1}, the WKB approach can still provide a reliable approximation for the dominant quasinormal frequencies.

The first-order WKB formula corresponds to the eikonal approximation, which becomes exact in the limit $\ell \to \infty$. The general WKB expression for quasinormal frequencies can be represented as an expansion around the eikonal limit \cite{Konoplya:2019hlu}:
\begin{eqnarray}\label{WKBformula-spherical}
\omega^2 &=& V_0 + A_2(\K^2) + A_4(\K^2) + A_6(\K^2) + \ldots \\ \nonumber
&-& \imo \K \sqrt{-2V_2} \left( 1 + A_3(\K^2) + A_5(\K^2) + A_7(\K^2) + \ldots \right),
\end{eqnarray}
where the matching conditions for the quasinormal modes impose
\[
\K = n + \frac{1}{2}, \quad n = 0, 1, 2, \ldots,
\]
with $n$ being the overtone number. Here, $V_0$ is the value of the effective potential at its maximum, $V_2$ is the second derivative of the potential at this point, and $A_i$ for $i = 2, 3, 4, \ldots$ are the $i$-th order WKB correction terms beyond the eikonal approximation. These corrections depend on $\K$ and the derivatives of the potential at its maximum up to the order $2i$. Explicit forms for the corrections $A_i$ can be found in \cite{Iyer:1986np} for the second and third WKB orders, in \cite{Konoplya:2003ii} for the 4th-6th orders, and in \cite{Matyjasek:2017psv} for the 7th-13th orders.
If the differences between the results obtained at consecutive WKB orders, such as the 5th, 6th, and 7th, are relatively small, it indicates that the WKB method is stable. In such cases, the relative error is typically of the same magnitude as these differences or smaller. Here we present the data obtained mostly by the 10th WKB order with Padé approximants \cite{Matyjasek:2017psv}, where the WKB series demonstrates a kind of plateau of ``relative convergence''.

\subsection{Time-domain integration}

The Gundlach-Price-Pullin time-domain integration scheme \cite{Gundlach:1993tp} is a widely used numerical method for analyzing the evolution of perturbations in black-hole spacetimes and extracting quasinormal modes. This approach is based on solving the wave equation, typically written in the form:
\[
\frac{\partial^2 \Psi}{\partial t^2} - \frac{\partial^2 \Psi}{\partial r_*^2} + V(r_*)\Psi = 0,
\]
where $\Psi$ is the perturbation field, $r_*$ is the tortoise coordinate, and $V(r_*)$ is the effective potential. The method discretizes this wave equation on a numerical grid in characteristic coordinates $u = t - r_*$ and $v = t + r_*$, transforming the equation into a form suitable for finite difference integration:
\begin{eqnarray}
\Psi\left(N\right)&=&\Psi\left(W\right)+\Psi\left(E\right)-\Psi\left(S\right)
-\Delta^2V\left(S\right)\frac{\Psi\left(W\right)+\Psi\left(E\right)}{8}+{\cal O}\left(\Delta^4\right),\label{Discretization}
\end{eqnarray}
Here, $N\equiv\left(u+\Delta,v+\Delta\right)$, $W\equiv\left(u+\Delta,v\right)$, $E\equiv\left(u,v+\Delta\right)$, and $S\equiv\left(u,v\right)$ represent neighboring points on the computational grid. This formulation avoids coordinate singularities near the event horizon, ensuring numerical stability and allowing for the accurate evolution of perturbations over time. This scheme captures the full time-domain evolution, including the initial burst, the quasinormal ringing phase, and the late-time tails.

Once the time-domain profiles of the perturbations are obtained, the Prony method is employed to extract the quasinormal frequencies. The method fits the time-domain signal $\Psi(t)$ to a sum of exponentially damped sinusoids:
\[
\Psi(t) = \sum_{i=1}^N A_i e^{-\imo \omega_i t},
\]
where $\omega_i$ represents the complex frequency. By solving the least squares problem for the coefficients $A_i$ and frequencies $\omega_i$, the contribution of the dominant modes can be isolated and analyzed. The combination of the Gundlach-Price-Pullin integration scheme and the Prony method provides a robust approach for determining quasinormal frequencies, offering insights into the stability and dynamics of perturbations in various space-time geometries.

\subsection{Bernstein polynomial method}

The Bernstein polynomial method offers a numerical approach to approximate the solutions of differential equations by expanding the unknown function in terms of Bernstein polynomials. The calculation of the coefficients is further performed using the pseudospectral method with Chebyshev collocation grid points \cite{Dias:2010eu,Jansen:2017oag}. This approach is particularly useful for handling boundary value problems due to the properties of Bernstein polynomials in approximating continuous functions over finite intervals.

Following \cite{Fortuna:2020obg}, we introduce the compact coordinate,
\begin{equation}\label{compact}
u\equiv\frac{\frac{1}{r}-\frac{1}{r_c}}{\frac{1}{r_h}-\frac{1}{r_c}},
\end{equation}
where $r_h$ is the event horizon of the black hole and $r_c$ is the cosmological horizon. For the asymptotically flat spacetime $r_c\to\infty$. For the traversable wormhole $r_h$ corresponds to the de Sitter horizon in the universe beyond the throat. So that for all the object under consideration $0\leq u\leq 1$.

We introduce the function $\psi(u)$, which is regular at $u=0$ and $u=1$ when $\omega$ is a quasinormal mode,
\begin{equation}\label{regularized}
\Psi(u)=F(u,\omega)\psi(u),
\end{equation}
where the prefactor $F(u,\omega)$ is defined from the characteristic equations at the singular points of the wavelike equation~(\ref{wave-equation}).

We approximate the solution $\psi(u)$ using Bernstein polynomials over the interval $[0, 1]$:
\begin{equation}\label{Bernsteinsum}
\psi(u) \approx \sum_{k=0}^{N} c_k \, B_{k,N}(u),
\end{equation}
where $c_k$ are coefficients to be determined, $N$ is the degree of the polynomial, and $B_{k,N}(u)$ are the Bernstein basis polynomials defined as:
\begin{equation}
B_{k,N}(u) = \frac{N!}{k!(N-k)!} \, u^k (1 - u)^{N - k}.
\end{equation}
Then we substitute the Bernstein expansion into the perturbation equation and use a Chebyshev collocation grid of $N+1$ points,
$$u_p=\frac{1}{2}\left(1-\cos \frac{p\cdot\pi}{N}\right)=\sin^2\frac{p\cdot\pi}{2N}, \qquad p=\overline{0,N}.$$
Thus, the problem is reduced to a set of linear equations with respect to $c_k$, which has nontrivial solutions iff the corresponding coefficient matrix is singular. The elements of the coefficient matrix are polynomials of $\omega$, therefore, we solve numerically the eigenvalue problem for a matrix pencil with respect to the quasinormal frequencies $\omega$, which is a linear problem.

In order to exclude the spurious eigenvalues, which appear due to finiteness of the polynomial basis in~(\ref{Bernsteinsum}), we compare both the eigenfrequencies and corresponding approximating polynomials for different values of $N$, as proposed in~\cite{Konoplya:2022xid}. Namely, from each set of the solutions we take the eigenvalues that differ less than the required accuracy and, for each pair of the corresponding eigenfunction, $\psi^{(1)}$ and $\psi^{(2)}$, we calculate
$$1-\frac{|\langle \psi^{(1)}\;|\;\psi^{(2)} \rangle|^2}{||\psi^{(1)}||^2||\psi^{(2)}||^2}=\sin^2\alpha,$$
where $\alpha$ is the angle between the vectors $\psi^{(1)}$ and $\psi^{(2)}$ in the $L^2$-space. If all values of $\alpha$ are sufficiently small, we conclude that the obtained eigenvalues $\omega$ approximate the quasinormal frequencies, and better approximations correspond to larger $N$. We estimate the error by calculating the difference between the approximate eigenvalues of $\omega$, corresponding to different values of $N$.

Bernstein polynomials possess excellent convergence properties for approximating continuous functions, which enhances the accuracy of the method. The method avoids issues related to divergent behaviors at the boundaries by working within a finite interval and using the appropriate boundary conditions. More details can be found in \cite{Fortuna:2020obg,Konoplya:2022xid,Konoplya:2022zav}.

\begin{figure*}
\resizebox{\linewidth}{!}{\includegraphics{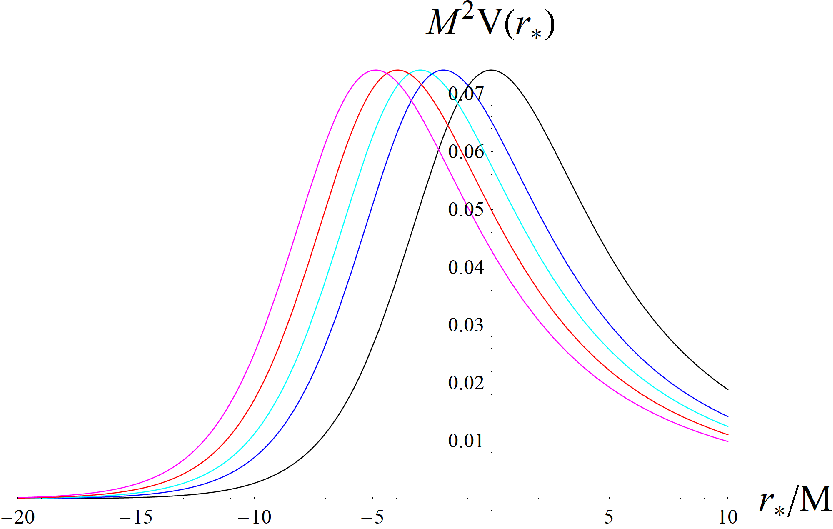}\includegraphics{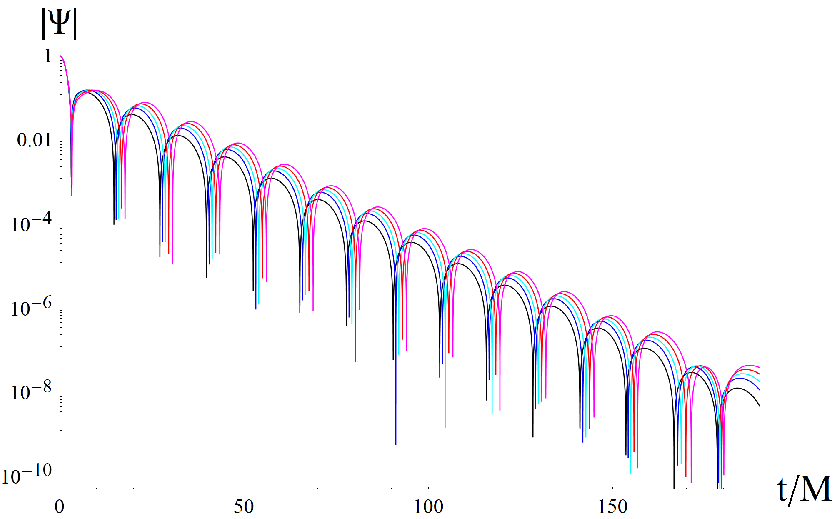}\includegraphics{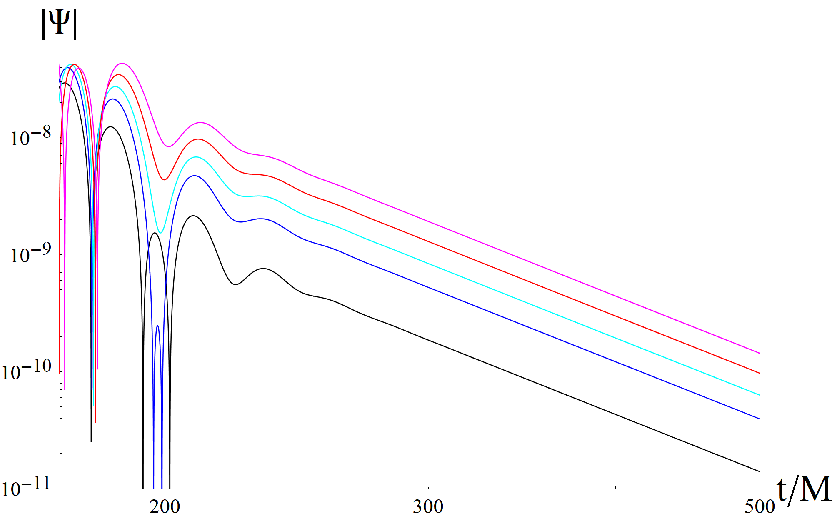}}
\caption{The effective potentials (left panel) and time-domain profiles (semi-log plot for the ringdown and log-log plot for the late-time tails) for the Maxwell perturbations ($\ell=1$) of the asymptotically flat Jizba-Mudruňka solution ($k=0$, $\gamma=0$, $\alpha=0$, $M=\beta$): $r_0=0$ (black), $r_0=3M$ (blue), $r_0=4M$ (cyan), $r_0=5M$ (red), $r_0=6M$ (magenta). $r_0=0$ corresponds to the Schwarzschild black hole, for $0<r_0\leq\pi M$ there is a black hole, which differs from Schwarzschild, when $\pi M<r_0<2\pi M$ there is a throat, which connects the flat space to a de Sitter universe. The time-domain profile is given for $r=3M$ ($r_*=0$).}\label{fig:flats1l1}
\end{figure*}

\begin{figure*}
\resizebox{\linewidth}{!}{\includegraphics{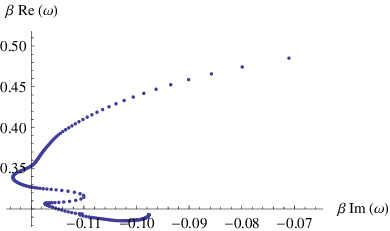}\includegraphics{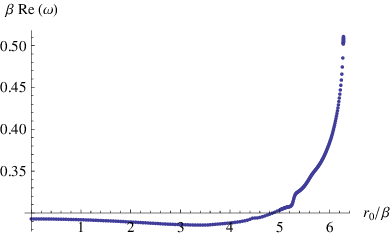}\includegraphics{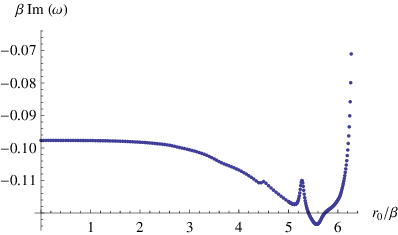}}
\caption{Scalar field perturbations ($\ell=1$, $n=0$) for $\gamma=0$ and $k=0$ ($M=\beta$, $\alpha=0$), for various values of $r_{0}$.}\label{fig:WeylL1gamma0}
\end{figure*}

\section{Quasinormal modes}\label{sec:QNMs}
\begin{table*}
\centerline{
\begin{tabular}{l c c c c}
\hline
\hline
$\gamma$ & $\omega$ ($r_{0} =0$) & $\omega$ ($r_{0} =3.14$) & $\omega$ ($r_{0} =5.5$) & $\omega$ ($r_{0} =6.28$) \\
\hline
 ~0 & 0.292936-0.097660 i     & 0.285841-0.101134 i  & 0.332508-0.122606 i  & 0.501895-0.051487 i \\
 -0.02 & 0.281714-0.095430 i  & 0.274984-0.098594 i  & 0.317741-0.117127 i  & 0.431138-0.094564 i \\
 -0.04 & 0.270344-0.092994 i  & 0.263994-0.095850 i  & 0.301728-0.108877 i  & 0.391830-0.101732 i \\
 -0.06 & 0.258802-0.090340 i  & 0.252844-0.092887 i  & 0.281835-0.105151 i  & 0.362062-0.100922 i \\
 -0.08 & 0.247061-0.087457 i  & 0.241502-0.089689 i  & 0.266269-0.103320 i  & 0.333921-0.101787 i \\
 -0.1 & 0.235089-0.084328 i   & 0.229939-0.086229 i  & 0.251729-0.099476 i  & 0.309310-0.099689 i \\
 -0.12 & 0.222846-0.080937 i  & 0.218138-0.082479 i  & 0.237268-0.094096 i  & 0.286487-0.096700 i \\
 -0.14 & 0.210284-0.077261 i  & 0.206090-0.078449 i  & 0.224242-0.088038 i  & 0.264066-0.093030 i \\
 -0.16 & 0.197342-0.073273 i  & 0.193704-0.074176 i  & 0.210587-0.084766 i  & 0.241196-0.087394 i \\
 -0.18 & 0.183940-0.068938 i  & 0.180844-0.069613 i  & 0.193964-0.080042 i  & 0.219571-0.080776 i \\
 -0.2 & 0.169970-0.064212 i   & 0.167396-0.064688 i  & 0.177228-0.073654 i  & 0.198729-0.074040 i \\
 -0.22 & 0.155281-0.059035 i  & 0.153215-0.059339 i  & 0.160483-0.066570 i  & 0.178081-0.068295 i \\
 -0.24 & 0.139648-0.053323 i  & 0.138071-0.053489 i  & 0.143303-0.059122 i  & 0.156284-0.062179 i \\
 -0.26 & 0.122703-0.046952 i  & 0.121583-0.047018 i  & 0.125028-0.051287 i  & 0.133666-0.054282 i \\
 -0.28 & 0.103771-0.039695 i  & 0.103066-0.039704 i  & 0.104739-0.042483 i  & 0.110043-0.044799 i \\
 -0.3 & 0.081407-0.031055 i   & 0.081056-0.031043 i  & 0.081794-0.032387 i  & 0.084300-0.033924 i \\
 -0.32 & 0.051130-0.019409 i  & 0.051042-0.019402 i  & 0.051205-0.019688 i  & 0.051694-0.020230 i \\
 -0.33 & 0.025484-0.009643 i  & 0.025473-0.009642 i  & 0.025495-0.009678 i  & 0.025552-0.009733 i \\
 -0.331 & 0.021315-0.008063 i & 0.021309-0.008062 i  & 0.021321-0.008083 i  & 0.021355-0.008116 i \\
\hline
\hline
\end{tabular}
}
\caption{Fundamental quasinormal mode for $\ell=1$ scalar field perturbations found by the 10th order WKB method with Padé approximants; $k=0$, $\beta =1$. The cases $r_{0} =0$ is the Mannheim-Kazanas BH, $r_{0} =3.14$ Jizba-Mudruňka BH near the transition with the wormhole, $r_{0} =5.5$ Jizba-Mudruňka wormhole, $r_{0} =6.28$ Jizba-Mudruňka wormhole near the transition with the asymptotic AdS space.}
\end{table*}

\begin{table*}
\centerline{
\begin{tabular}{l c c c}
\hline
\hline
$\gamma$ & $\omega$ ($\ell =1$) & $\omega$ ($\ell =2$)  & $\omega$ ($\ell =3$)  \\
\hline
 ~0 & 0.248263-0.092488 i     & 0.457596-0.095004 i  & 0.656899-0.095616 i \\
 -0.02 & 0.240719-0.089675 i  & 0.443666-0.092112 i  & 0.636895-0.092704 i \\
 -0.04 & 0.232964-0.086781 i  & 0.429305-0.089128 i  & 0.616255-0.089699 i \\
 -0.06 & 0.224973-0.083795 i  & 0.414463-0.086044 i  & 0.594913-0.086591 i \\
 -0.08 & 0.216711-0.080706 i  & 0.399085-0.082848 i  & 0.572786-0.083369 i \\
 -0.1 & 0.208141-0.077502 i   & 0.383100-0.079525 i  & 0.549776-0.080018 i \\
 -0.12 & 0.199213-0.074166 i  & 0.366424-0.076060 i  & 0.525765-0.076521 i \\
 -0.14 & 0.189868-0.070678 i  & 0.348954-0.072431 i  & 0.500604-0.072858 i \\
 -0.16 & 0.180029-0.067013 i  & 0.330555-0.068611 i  & 0.474105-0.069001 i \\
 -0.18 & 0.169598-0.063136 i  & 0.311057-0.064566 i  & 0.446026-0.064915 i \\
 -0.2 & 0.158440-0.059001 i   & 0.290229-0.060249 i & 0.416039-0.060554 i \\
 -0.22 & 0.146373-0.054543 i  & 0.267751-0.055592 i  & 0.383691-0.055851 i \\
 -0.24 & 0.133127-0.049661 i  & 0.243155-0.050501 i  & 0.348319-0.050709 i \\
 -0.26 & 0.118286-0.044196 i  & 0.215704-0.044820 i  & 0.308873-0.044976 i \\
 -0.28 & 0.101128-0.037869 i  & 0.184112-0.038280 i  & 0.263522-0.038384 i \\
 -0.3 & 0.080156-0.030103 i   & 0.145688-0.030317 i  & 0.208431-0.030372 i \\
 -0.32 & 0.050828-0.019158 i  & 0.092232-0.019214 i  & 0.131890-0.019230 i  \\
 -0.33 & 0.025448-0.009611 i  & 0.046140-0.009619 i &   0.065963-0.009620 i \\
 -0.331 & 0.021294-0.008044 i & 0.038605-0.008048 i  &  0.055190-0.008050 i \\
\hline
\hline
\end{tabular}
}
\caption{Fundamental quasinormal mode for $\ell=1,2,3$ electromagnetic field perturbations found by the 10th order WKB method with Padé approximants; $k=0$, $\beta=1$.}
\end{table*}

\begin{table*}
\centerline{
\begin{tabular}{l c c c}
\hline
\hline
$\gamma$ & $\omega$ ($\ell =1/2$) & $\omega$ ($\ell =3/2$) & $\omega$ ($\ell =5/2$) \\
\hline
 ~0 & 0.182643-0.096566 i     & 0.380041-0.096408 i & 0.574094-0.096305 i \\
 -0.02 & 0.177102-0.093614 i  & 0.368476-0.093468 i & 0.556614-0.093370 i \\
 -0.04 & 0.171428-0.090546 i  & 0.356559-0.090427 i & 0.538582-0.090337 i \\
 -0.06 & 0.165598-0.087355 i  & 0.344250-0.087275 i & 0.519939-0.087196 i \\
 -0.08 & 0.159589-0.084030 i  & 0.331502-0.084003 i & 0.500614-0.083937 i \\
 -0.1 & 0.153373-0.080561 i   & 0.318256-0.080595 i & 0.480521-0.080545 i \\
 -0.12 & 0.146913-0.076935 i  & 0.304442-0.077037 i & 0.459555-0.077004 i \\
 -0.14 & 0.140166-0.073137 i  & 0.289974-0.073309 i & 0.437587-0.073293 i \\
 -0.16 & 0.133072-0.069147 i  & 0.274739-0.069385 i & 0.414451-0.069386 i \\
 -0.18 & 0.125557-0.064937 i  & 0.258594-0.065231 i & 0.389935-0.065248 i \\
 -0.2 & 0.117518-0.060470 i   & 0.241345-0.060804 i & 0.363753-0.060834 i \\
 -0.22 & 0.108811-0.055691 i  & 0.222723-0.056038 i & 0.335505-0.056077 i \\
 -0.24 & 0.099222-0.050513 i  & 0.202336-0.050840 i & 0.304611-0.050883 i \\
 -0.26 & 0.088420-0.044787 i  & 0.179567-0.045059 i & 0.270152-0.045101 i \\
 -0.28 & 0.075834-0.038240 i  & 0.153337-0.038430 i & 0.230521-0.038463 i \\
 -0.3 & 0.060303-0.030295 i   & 0.121398-0.030391 i & 0.182359-0.030412 i \\
 -0.32 & 0.038355-0.019210 i  & 0.076897-0.019233 i & 0.115414-0.019240 i  \\
-0.33 & 0.019226-0.009620 i  & 0.038479-0.009621 i & 0.057728-0.009622 i  \\
-0.331 & 0.016090-0.008050 i  & 0.032197-0.008050 i & 0.048300-0.008050 i  \\
\hline
\hline
\end{tabular}
}
\caption{Fundamental quasinormal mode ($n=0$) for Dirac field perturbations found by the 6th order WKB method with Padé approximants; $k=0$, $\beta =1$.}
\end{table*}

The distinctive feature of the quasinormal spectrum of black holes and wormholes in the Weyl gravity is the presence of three branches of modes:
\begin{enumerate}
    \item Schwarzschild branch of modes, which reduces to the well-known quasinormal modes of Schwarzschild spacetime when $\alpha =\gamma =k =0$. In other words, these are modes of Schwarzschild black holes corrected by the dark matter term $\gamma$ and cosmological constant $k$.
    \item Dark-matter-induced branch of quasinormal modes: When $k=0$ and the cosmological horizon appears due to the effective dark-matter term $\gamma<0$, as the mass of the black hole goes to zero, these modes tend to modes of empty spacetime in Weyl gravity, which are similar in nature to those in the de Sitter space.
    \item de Sitter branch of frequencies: When $\gamma =0$ and the mass vanishes, these modes go over into the modes of pure de Sitter spacetime \cite{Lopez-Ortega:2012xvr,Lopez-Ortega:2007vlo}. Thus, these modes are eigenvalues of empty de Sitter space corrected by the presence of a black hole.
\end{enumerate}

All these branches are evident for the Mannheim-Kazanas solution \cite{Mannheim:1988dj}, but not for the black holes and wormholes found in \cite{Jizba:2024owd}, as the limit $M\to0$ is not well-defined for those solutions.

\begin{figure*}
\resizebox{\linewidth}{!}{\includegraphics{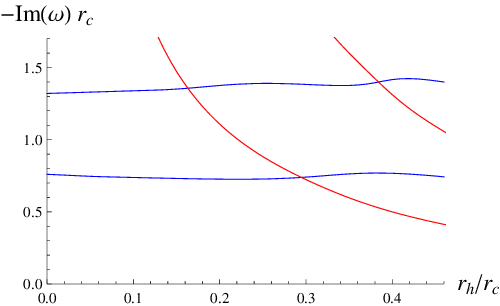}\includegraphics{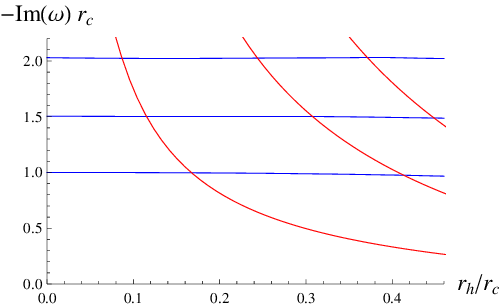}}
\caption{Imaginary part of the cosmological branch (blue) and black-hole branch (red) for the Mannheim-Kazanas solution ($r_0=0$) in the limit of vanishing effective cosmological constant ($k=0$) for $s=\ell=0$ (left panel) and $s=\ell=1$ (right panel): As black-hole radius $r_h\to0$ ($M\to0$) the cosmological branch approaches the analytic solution, (\ref{MKzeroEM}) and (\ref{MKzeroS}) for the electromagnetic and scalar perturbations respectively, while the black-hole branch ($\omega\propto r_h^{-1}$) diverges. The quasinormal frequencies were calculated using the Bernstein polynomial method.}\label{fig:qnmlimit}
\end{figure*}

The Schwarzschild branch of modes is determined with remarkable accuracy using the WKB method, provided $\ell$ is not smaller than $n$. The longest-lived mode of this branch, corresponding to the fundamental mode of the Schwarzschild solution, is presented in Tables 1–3. However, the longest-lived modes of the other two branches can only be detected via time-domain integration, while the overtones require the application of the Bernstein polynomial method. From Fig.~\ref{fig:qnmlimit} we can see that the purely imaginary, i.e. non-oscillatory, quasinormal modes of the dark matter branch exist in the spectrum. When the analog of the de Sitter radius, determining the cosmological horizon owing to the effective dark matter term, is much larger than the black hole radius, these modes go over into the modes of empty spacetime in the Weyl gravity, given by Eqs.~(\ref{MKzeroEM}) and~(\ref{MKzeroS}).

The modes associated with the Schwarzschild branch converge to a universal limit as the effective dark matter parameter $\gamma$ approaches its extreme value. This limit is independent of both $r_0$ and the spin of the field under consideration, as illustrated in Tables 1–3. When mass of the black hole goes to zero these modes, being inverse proportional to it, diverges as shown in Fig.~\ref{fig:qnmlimit}. When the dark matter term goes to zero these modes go over into the Schwarzschild ones.

We also observe that the spectrum of wormholes with large $r_0$ differs significantly from that of black holes, exhibiting much larger real oscillation frequencies, as shown in Tables 1–3. As demonstrated in Fig.~\ref{fig:WeylL1gamma0}, the transition between a black hole and a wormhole, which occurs at $r_0 = \pi\beta$, is not particularly remarkable. Quasinormal modes at the black hole–wormhole transition have been studied in several works \cite{Churilova:2019cyt,Bronnikov:2019sbx}. Typically, this transition is characterized by the onset of echoes once the wormhole is formed due to the appearance of a double peak in the effective potential. Here, however, no such doubling of the potential peak occurs during the transition to the wormhole state, and consequently, no echoes are observed.

The only notable feature of the black-hole-wormhole transition point is that the minimum of the real oscillation frequency occurs at $r_0\approx3\beta$ (see Fig.~\ref{fig:WeylL1gamma0}).

In the eikonal limit ($\ell\to\infty$), the effective potential does not depend on the spin,
\begin{equation}
    V(r)=\left(\ell+\frac{1}{2}\right)^2\left(P(\rho(r))+\Order{\frac{1}{\ell}}\right),
\end{equation}
and its point of maximum is reached (cf.~\ref{Omegac}) for
\begin{equation}
\rho(r_m)=3\beta+\Order{\frac{1}{\ell}}.
\end{equation}
Therefore, from (\ref{rhodef}), we can obtain analytic expression for the position of the peak of the potential barrier
\begin{equation}
r_{ m} = \frac{r_0}{\tan{(r_0/3\beta)}}+\Order{\frac{1}{\ell}},
\end{equation}
and, consequently for quasinormal modes using the expansion in terms of the inverse multiple number in the first order WKB formula

\begin{equation}\label{eikonalQNMs}
\omega=\frac{\sqrt{1+3\beta\gamma-27k\beta^2}}{3\sqrt{3}\beta}\left(\ell+\frac{1}{2}-\imo\left(n+\frac{1}{2}\right)+\Order{\frac{1}{\ell}}\right),
\end{equation}
which generalizes the formula (19) of \cite{Zhidenko:2003wq}. The latter can be obtained from (\ref{eikonalQNMs}) for $\gamma=0$, $\beta=M$, and $k=\Lambda/3$.

The quasinormal modes in the eikonal limit (\ref{eikonalQNMs}) can be represented through the shadow radius (\ref{shadowdef}) and Lyapunov exponent (\ref{Lyapunovdef}) as follows:
\begin{equation}\label{eikonalformula}
\omega=\frac{1}{R_{s}}\left(\ell+\frac{1}{2}\right)-\imo\lambda\left(n+\frac{1}{2}\right)+\Order{\frac{1}{\ell}}.
\end{equation}

Let us note that the above eikonal formula applies only to the Schwarzschild branch of modes and does not account for the de Sitter and dark matter branches. Consequently, the correspondence between the eikonal quasinormal modes and null geodesics or shadows \cite{Cardoso:2008bp,Jusufi:2019ltj} should be interpreted cautiously, keeping in mind that there are notable exceptions to this correspondence, particularly when higher curvature corrections or a cosmological constant are included \cite{Konoplya:2017wot,Konoplya:2020bxa,Konoplya:2022gjp,Bolokhov:2023dxq}. Therefore, it remains uncertain whether this correspondence will hold for gravitational perturbations in the Weyl theory, which lie beyond the scope of this investigation.

In the common range of applicability, we observe excellent agreement among all the methods. For instance, for $\ell = 1$ electromagnetic perturbations shown in Fig.~3, the time-domain integration and extraction of frequencies using the Prony method yield, for the Schwarzschild branch, $\omega M = 0.248270 - 0.092488\imo$, while the WKB method gives $\omega M = 0.248263-0.092488\imo$. Thus, the relative difference between the frequencies obtained by these two methods remains within a small fraction of one percent in this case.


It turns out that grey-body factors of a wave when scattering near the peak of the potential barrier might be more stable characteristic linked to the gravitational wave profile \cite{Oshita:2021iyn,Rosato:2024arw}. Using the numerical data for the fundamental mode ($\omega_0$) and the first overtone ($\omega_1$) for black holes and wormholes one can obtain the transmission coefficients or grey-body factors via the correspondence established in \cite{Konoplya:2024lir} for black holes and in \cite{Bolokhov:2024otn} for wormholes:
\begin{equation}\label{transmission-eikonal}
\Gamma_{\ell}(\Omega)=\left(1+e^{2\pi\dfrac{\Omega^2-\re{\omega_0}^2}{4\re{\omega_0}\im{\omega_0}}}\right)^{-1} + \Sigma(\omega_{0},\omega_{1}).
\end{equation}
Here $\Sigma(\omega_{0},\omega_{1})$ is the sum of the correction terms beyond the eikonal limit found in \cite{Konoplya:2024lir}. This correspondence has been recently tested in \cite{Malik:2024cgb,Skvortsova:2024msa,Dubinsky:2024vbn}.

\begin{figure*}
\resizebox{\linewidth}{!}{\includegraphics{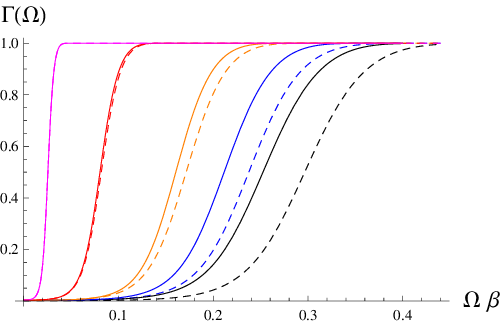}\includegraphics{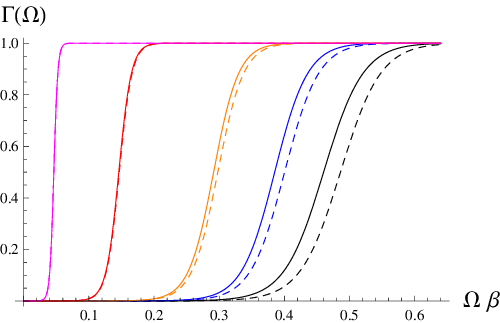}}
\caption{Grey-body factors of the Mannheim-Kazanas solution ($k=0$) calculated using the two dominant quasinormal modes via the correspondence formula for $\gamma=0$ (Schwarzschild, black), $\gamma\beta=0.1$ (blue), $\gamma\beta=0.2$ (orange), $\gamma\beta=0.3$ (red), and $\gamma\beta=0.33$ (magenta, left) for the scalar (dashed line) and electromagnetic (solid line) fields: $\ell=1$ (left panel) and $\ell=2$ (right panel).}\label{fig:gbfactors}
\end{figure*}

In Fig.~\ref{fig:gbfactors}, we present the grey-body factors for the scalar and electromagnetic fields of the Mannheim-Kazanas solution ($k=0$) for different values of the dark matter parameter $\gamma$. A key observation is that, as $\gamma$ approaches its extreme value, the grey-body factors for integer-spin fields converge to a universal limit, independent of the spin of the field. This behavior closely resembles the corresponding feature observed in the quasinormal mode spectrum. Additionally, it is noteworthy that the quasinormal modes for electromagnetic and Dirac fields depend on the black hole or wormhole mass $M$ only through the parameter $\beta$ and are independent of the parameter $r_0$. Consequently, since the mass of the compact object is typically not known with sufficient accuracy, and the quality factor \(Q \propto |\re{\omega}/\im{\omega}|\) is frequently used for comparison with observations, the quality factor for black holes and wormholes with different values of $r_{0}$ will be indistinguishable.

\section{Conclusions}\label{sec:conclusions}

The Weyl gravity suggests solutions to three interesting problems:
(a) dark matter problem, without introducing dark matter fields,
(b) reproducing the cosmological constant without formal introduction of it in the Lagrangian,
(c) existence of traversable wormholes. The black-hole solution is not unique in this theory: For each value of the cosmological constant, the dark-matter term, and the asymptotic mass there is a one-parametric family of black-hole solutions obtained in \cite{Jizba:2024owd}, approaching the solutions of \cite{Mannheim:1988dj} when the additional parameter $r_0$ goes to zero. Moreover, the plethora of solutions in the Weyl theory include asymptotically de Sitter-like and anti-de Sitter black holes and wormholes.

In the present paper we have studied quasinormal spectrum of asymptotically de Sitter-like black holes and wormholes in the Weyl gravity. It was shown that the spectrum consists of the three branches of modes which dominate at different periods of time: Schwarzschild branch, dark matter branch and de Sitter branch. However, the limit of vanishing black hole mass is well defined only for the Mannheim-Kazanas black hole solution \cite{Mannheim:1988dj} and for this case we show that the modes of a black holes goes to those of the empty spacetime when mass goes to zero. The perturbation equations for the empty space in the Weyl gravity allow for an exact solution. Consequently, the quasinormal modes of the empty space in the Weyl gravity are obtained in the analytic form.

Quasinormal modes depend essentially on the parameters of the solutions responsible for the effective dark matter and cosmological terms, as well as on the parameter $r_0$ which distinguishes the three compact objects (Schwarzschild-like and non-Schwarzschild-like black holes, and a wormhole).
In the eikonal limit we found the analytic expression for quasinormal frequencies, which respects the correspondence between quasinormal modes and null geodesics/radius of the shadow for all three branches.

In addition we have found the exact analytic expression for the radius of the shadow cast by the black hole or wormhole in the Weyl gravity.

Here we were limited by asymptotically de Sitter-like black holes. However, our analysis could in principle be extended to asymptotically anti-de Sitter spacetimes. For this case, quasinormal modes of vacuum spacetime is also expected to be the limit of black hole's frequencies, when the mass goes to zero \cite{Konoplya:2002zu}.

\acknowledgments
A.K. is grateful to the Excellence Project FoS UHK 2204/2025–2026 for the financial support.\\
A.Z. acknowledges the Research Centre for Theoretical Physics and Astrophysics at the Institute of Physics, Silesian University in Opava, for their hospitality.

\bibliographystyle{JHEP}
\bibliography{Bibliography}

\end{document}